\pdfoutput=1

\documentclass[11pt]{article}

\usepackage{naacl2021} 

\usepackage{times}
\usepackage{latexsym}

\usepackage[T1]{fontenc}

\usepackage[utf8]{inputenc}

\usepackage{microtype}
\usepackage{times}
\usepackage{latexsym}
\usepackage{lipsum}
\usepackage{booktabs}
\usepackage{array}
\usepackage{multirow}
\usepackage{graphicx}
\usepackage[normalem]{ulem}

\usepackage{etoolbox}
\usepackage{todonotes}
\usepackage{hyperref}
\usepackage{enumerate}
\usepackage{enumitem}
\usepackage{amsmath}
\usepackage{color}
\usepackage{tcolorbox}
\usepackage{CJK}
\usepackage{adjustbox}
\tcbset{width=0.9\textwidth,boxrule=0pt,colback=red,arc=0pt,auto outer arc,left=0pt,right=0pt,boxsep=5pt}
\usepackage[font=small,skip=-5pt]{subcaption}
\usepackage{diagbox}

\definecolor{citeworthy}{HTML}{32A852}

\usepackage{microtype}
\usepackage{comment}

\newcommand\dataset{\textsc{CiteWorth}}

\title{Determining the Credibility of Science Communication}

\author{Isabelle Augenstein \\
  Dpt. of Computer Science \\
  University of Copenhagen \\
  \texttt{augenstein@di.ku.dk}
  } 

\begin{document}
\maketitle
\begin{abstract}
Most work on scholarly document processing assumes that the information processed is trustworthy and factually correct. However, this is not always the case. There are two core challenges, which should be addressed: 1) ensuring that scientific publications are credible -- e.g. that claims are not made without supporting evidence, and that all relevant supporting evidence is provided; and 2) that scientific findings are not misrepresented, distorted or outright misreported when communicated by journalists or the general public. I will present some first steps towards addressing these problems and outline remaining challenges.
\end{abstract}

\section{The Life Cycle of Scientific Research}
Scientific research is highly diverse not just when it comes to the topic of study, but also how studies are conducted, how the resulting research is described and when and where it is published. However, what different fields still have in common is a certain life cycle, starting with planning a study and ending with promoting the research post-publication, in the hopes of the article finding readership and having an impact.

Scholary document processing aims to support researchers throughout this life cycle of scientific research, by offering various tools to automate otherwise manual processes.
Most research within scholarly document processing has focused on supporting information discovery for finding related work. Most prominently, research has focused on methods to condense scientific documents, using entity extraction and linking, keyphrase or relation extraction \cite{augenstein-etal-2017-semeval,augenstein-sogaard-2017-multi,conf/akbc/WrightKMH19,gabor2018semeval,ammar-etal-2018-construction} or automatic summarisation \cite{collins-etal-2017-supervised,yasunaga2019scisummnet}.
 
Once papers are written and submitted for peer review, it is pertinent to evaluate them fairly and objectively. This process is far from straight-forward, as, among others, reviewers have certain biases, including against truly novel research \cite{rogers-augenstein-2020-improve,bhattacharya2020stagnation}.
Research has thus focused on automatically generating peer reviews from paper content \cite{wang-etal-2020-reviewrobot}, as well as on studying how well review scores can be predicted from review texts \cite{kang-etal-2018-dataset,plank2019citetracked}.

Finally, post-publication, the impact of scientific work can be tracked, using citations and citation counts as a proxy for this. It is again worth noting that there are significant biases in this -- e.g. author information is among the, if not the most salient feature for predicting citation counts \cite{yan2011citation,holm2020longitudinal}. 
Looking further into what papers are cited and why, \citet{mohammad-2020-gender,mohammad-2020-examining} find that there are significant topical as well as gender biases when it comes to who is cited and by whom.


\begin{table*}[!htp]
    \centering
    \fontsize{10}{10}\selectfont
    \begin{tabular}{p{15cm}}
    \toprule 
    \textbf{Biology} \\
    \midrule
    Wood Frogs (Rana sylvatica) are a charismatic species of frog common in much of North America. They breed in explosive choruses over a few nights in late winter to early spring. \textcolor{citeworthy}{\emph{The incidence in Wood Frogs was associated with a die-off of frogs during the breeding chorus in the Sylamore District of the Ozark National Forest in Arkansas\sout{ (Trauth et al., 2000)}.}} \\
    \bottomrule \\
    \toprule
    \textbf{Computer Science} \\
    \midrule
    \textcolor{citeworthy}{\emph{Land use or cover change is a direct reflection of human activity, such as land use, urban expansion, and architectural planning, on the earth's surface caused by urbanization\sout{ [1].}}} Remote sensing images are important data sources that can efficiently detect land changes. \textcolor{citeworthy}{\emph{Meanwhile, remote sensing image-based change detection is the change identification of surficial objects or        geographic phenomena through the remote observation of two or more different phases\sout{ [2]}.}} \\
    \bottomrule 

    \end{tabular}
    \caption{Excerpts from training samples in \dataset\ \cite{wright-augenstein-2021-citeworth} from the Biology and Computer Science fields. Green sentences are cite-worthy sentences, from which citation markers are removed during dataset construction.}
    \label{tab:dataset_examples}
\end{table*}

\section{Credibility and Veracity of Science Communication}

While all of the work referenced above is important in supporting researchers, it neglects one crucial aspect, namely that it assumes the resulting scientific documents and broader communication about them are credible and supported by the underlying evidence. Though it is the task of peer reviewers to spot issues regarding credibility, and the task of journalists to check their sources when they report on scientific studies, distortions, exaggerations and outright misrepresentations can still happen. 

The ongoing COVID-19 pandemic has highlighted the disastrous and direct consequences misreporting of scientific findings can have on our everyday lives, yet, there is still relatively little work on detecting issues in the credibility of scientific writing. This especially holds for detecting smaller nuances of untrustworthy scientific writing, whereas there is comparatively more work on detecting outright scientific misinformation \cite{vijjali-etal-2020-two,lima2021university}. 

Here, we highlight two important and so far understudied tasks to address issues with such smaller nuances of untrustworthy scientific writing, which can come into play at different stages of the life cycle of scientific research.
The first one is \textit{cite-worthiness detection}, which is about detecting whether or not a sentence ought to contain a citation to prior work. This task could help to ensure that claims are not made without supporting evidence, i.e. support researchers in writing more trustworthy scientific publications.

The second task is \textit{exaggeration detection}, which is to determine whether a statement describing the findings of a scientific study exaggerates them, e.g. by claiming that two variables are strongly correlated when in reality they only co-occur.
We argue that this task could be useful to verify if popular science reporting faithfully describes scientific research, or also to determine whether citation sentences (sentences which contain a citation; also called \textit{citances}) faithfully describe the research documented in the cited papers.

\begin{table*}[!htp]
    \centering
    \fontsize{10}{10}\selectfont
    \begin{tabular}{p{15cm}}
    \toprule 
    \textbf{Exaggerated Claims} \\
    \midrule
    \textbf{Press Release:} Players of the game rock paper scissors subconsciously copy each other’s hand shapes, significantly increasing the chance of the game ending in a draw, according to new research. \\
    \\
    \textbf{Abstract:} Specifically, the execution of either a rock or scissors gesture by the blind player was predictive of an imitative response by the sighted player. \\
    \bottomrule \\
    \toprule
    \textbf{Exaggerated Advice} \\
    \midrule
    \textbf{Press Release:} Parents should dilute fruit juice with water or opt for unsweetened juices, and only allow these drinks during meals.
     \\
     \\
    \textbf{Abstract:} Manufacturers must stop adding unnecessary sugars and calories to their FJJDS.
    \\
    \bottomrule 

    \end{tabular}
    \caption{Examples of exaggerated claims and exaggerated advice given in press releases about scientific papers.}
    \label{tab:exaggeration_examples}
\end{table*}

\subsection{Cite-Worthiness Detection}

\paragraph{The \dataset\ Dataset} To study cite-worthiness detection, we first introduce a new rigorously curated dataset, \dataset\ \cite{wright-augenstein-2021-citeworth}, for cite-worthiness detection from scientific articles. It is created from S2ORC, the Semantic Scholar Open Research Corpus \cite{lo2020s2orc}. \dataset\ consists of 1.2M sentences, balanced across 10 diverse scientific fields. While others have studied this task for few and/or narrow domains \cite{sugiyama2010identifying,farber2018cite}, and have also studied very related tasks, such as claim check-worthiness detection \cite{wright-augenstein-2020-claim} or citation recommendation \cite{jurgens2018measuring}, this is the largest and most diverse dataset for this task to date.  

An excerpt of our introduced dataset, \dataset\, can be found in Table \ref{tab:dataset_examples}. The dataset curation process involves: 1) data filtering, to identify credible papers with relevant metadata such as venue information; 2) citation span identification and masking, of which we only keep papers with citation spans at the end of sentences to avoid rendering sentences ungrammatical; 3) discarding paragraphs without citations, or where not all sentences have citation spans in accordance with our heuristics; 4) evenly sampling paragraphs, such that the resulting dataset is equally balanced for the domains of Biology, Medicine, Engineering, Chemistry, Psychology, Computer Science, Materials Science, Economics, Mathematics, and Physics.

Given this dataset, we then study: how cite-worthy sentences can be detected automatically; to what degree there are domain shifts between how different fields use citations; and if cite-worthiness data can be used to perform transfer learning to downstream scientific text tasks.

\paragraph{Methods for Cite-Worthiness Detection}
We find that the best performance can be achieved by a Longformer-based model \cite{DBLP:journals/corr/abs-2004-05150}, which encodes entire paragraphs in papers and jointly predicts cite-worthiness labels for each of the sentences contained in the paragraph. Additional gains in recall can be achieved by using positive unlabelled learning, as documented in \citet{wright-augenstein-2020-claim} for the related task of claim check-worthiness detection. Our best-performing model outperforms baselines such as a carefully fine-tuned SciBERT \cite{beltagy2019scibert} by over 5 points in F1.

\paragraph{Domain Differences}
To study domain effects, we perform a cross-evaluation, where we hold out one domain for testing and evaluate model performance on that, and compare this against an in-domain evaluation setting, where all domains observed at test time are also observed at training time.
We find that there is a high variance in the maximum performance for each field ($\sigma$ = 3.32), and between different fields on the same test data, despite large pretrained Transformer models being relatively invariant across domains~\cite{wright-augenstein-2020-transformer}. This suggests stark differences in how different fields employ citations.

\paragraph{Downstream Applicability}
We evaluate our models on downstream scientific document processing tasks from \citet{beltagy2019scibert}, which can be grouped into: named entity recognition tasks; relation extraction tasks; and text classification tasks.
Specifically, we use our best-performing model, pre-trained for cite-worthiness detection and masked language modelling, and fine-tune them for 10 different downstream tasks. We find that improvements over the state of the art can be achieved for two citation intent classification tasks.

\subsection{Exaggeration Detection}

We frame exaggeration detection in the context of popular science communication. Specifically, we ask the question: how can one automatically detect if popular science articles overstate the claims made in scientific articles?

Prior work has shown that exaggeration of findings of scientific articles is highly prevalent \cite{sumner2014association,bratton2019association,woloshin2009press,woloshin2002press}. Exaggeration can mean a sensationalised take-away of the applicability of the work in terms, i.e. giving advice for which there is no scientific basis. Moreover, the strength of the main causal claims and conclusions of a paper can be exaggerated. \autoref{tab:exaggeration_examples} shows examples of those two types of claims from the datasets curated by \citet{sumner2014association} and \citet{bratton2019association}, which we use in our work. 

Prior work \cite{yu2019detecting,yu2020measuring,li2017nlp} uses datasets based on PubMed abstracts and paired press releases from EurekAlert.\footnote{https://www.eurekalert.org/} Their core limitations of is that they are limited to only observational studies from PubMed, which have structured abstracts, which strongly simplifies the task of identifying the main claims of a paper. This also holds for the test settings they consider, meaning that the proposed models have a limited applicability.

By contrast, we study how to best identify exaggerated claims in popular science communication in the wild, without highly curated data with annotations about core claims. This represents a more realistic experimental setup, which is more suited to supporting downstream use cases such as flagging exaggerated popular news articles as well as exaggerated summaries of scientific papers as referenced in other scientific papers.

Our method is a semi-supervised approach, which first identifies sentences containing claims in both scientific articles and popular science communication within the medical domain,  
then identifies the main conclusion of both articles, 
and lastly predicts to what degree popular science articles exaggerate those findings.
We further analyse to what degree exaggeration of findings is correlated with the perceived media bias of popular science communication outlets.

\section{Conclusion}
This paper discusses research avenues for automatically determining the credibility of science communication, both in terms of scientific papers and popular science communication. These avenues are put in the context of scholarly data processing more broadly, and how different tasks can be used to assist the life cycle of scientific research. While existing research has focused on developing models for assisting with information discovery, peer review and citation tracking, comparatively little work has been done on identifying non-credible claims and assisting authors in making sure their research is backed up by sufficient evidence where needed. The suggestion is therefore to focus on two tasks: cite-worthiness detection, to identify sentences requiring citations; and exaggeration detection, to identify cases in which scientific findings have been overstated.
A core problem for both tasks is the lack of appropriate training data, which we address by introducing a new dataset, and a semi-supervised learning method, respectively. We hope our research will inspire future work on developing tools to assist authors and journalists in ensuring that research is described in a credible and evidence-based way.

\section*{Acknowledgements}

$\begin{array}{l}\includegraphics[width=1cm]{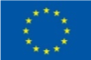} \end{array}$ The research documented in this paper has received funding from the European Union's Horizon 2020 research and innovation programme under the Marie Sk\l{}odowska-Curie grant agreement No 801199. 

Thank you to Dustin Wright for the fruitful discussions and feedback on this extended abstract.

\bibliography{anthology,custom}
\bibliographystyle{acl_natbib}




\end{document}